\documentclass{article}
\usepackage{spconf,amsmath}
\usepackage{amsthm}
\usepackage{amsfonts}
\usepackage{amssymb}
\usepackage{mathrsfs}
\usepackage{mathtools}
\usepackage[english]{babel}
\usepackage{float}
\usepackage[toc,page]{appendix}
\usepackage[pdftex]{graphicx}
\usepackage{epstopdf}
\usepackage{amsmath}
\usepackage{color}
\usepackage{multirow}
\usepackage{graphicx}
\usepackage{IEEEtrantools}
\usepackage{bm}
\usepackage{balance}
\usepackage{microtype}
\usepackage[caption=false,font=footnotesize]{subfig}

\usepackage{url}

\usepackage{cite}

\usepackage[]{algorithmicx}
\usepackage{algpseudocode,algorithm} 

\usepackage{color}
\usepackage[table]{xcolor}

\DeclareMathOperator*{\maximize}{maximize} 
\DeclareMathOperator*{\minimize}{minimize} 
\DeclareMathOperator*{\subjectto}{subject \hspace{3pt} to} 

\title{Low-Complexity Limited-Feedback Deep Hybrid Beamforming for Broadband Massive MIMO Communications}
\name{Ahmet~M.~Elbir$^1$ and Kumar Vijay~Mishra$^2$ 
}
\address{$^{1}$Department of Electrical and Electronics Engineering, D\"{u}zce University, D\"{u}zce 81620 Turkey\\
	$^{2}$The University of Iowa, Iowa City, IA 52242 USA 
	}

\begin{document}
	\ninept
    \setlength{\abovedisplayskip}{3pt}
    \setlength{\belowdisplayskip}{3pt}

	\maketitle
	\begin{abstract}
		In broadband millimeter-wave (mm-Wave) systems, it is desirable to design hybrid beamformers with common analog beamformer for the entire band while employing different baseband beamformers in different frequency sub-bands. Furthermore, the performance mostly relies on the perfectness of the channel information. In this paper, we propose a deep learning (DL) framework for hybrid beamformer design in broadband mm-Wave massive MIMO systems. We design a convolutional neural network (CNN) that accepts the channel matrix of all subcarriers as input and the output of CNN is the hybrid beamformers. The proposed CNN architecture is trained with imperfect channel matrices in order to provide robust performance against the deviations in the channel data. Hence, the proposed precoding scheme can handle the imperfect or limited feedback scenario where the full and exact knowledge of the channel is not available. We show that the proposed DL framework is more robust and computationally less complex than the conventional optimization and phase-extraction-based approaches.
	\end{abstract}
	%
	
	\section{Introduction}
	\label{sec:Introduciton}
	The conventional cellular communications systems suffer from spectrum shortage while the demand for wider bandwidth and higher data rates is continuously increasing \cite{mimoOverview}. In this context, millimeter wave (mm-Wave) band is a preferred candidate for fifth-generation (5G) communications technology \cite{5GwhatWillItBe} because they provide higher data rate and wider bandwidth \cite{mimoOverview,mishra2019toward}. As a consequence, they have become a leading candidate for the fifth generation (5G) wireless networks \cite{5GwhatWillItBe,hodge2019reconfigurable,ayyar2019robust}. The mm-Wave systems leverage large-scale antenna arrays to compensate for the propagation losses at high frequencies. Hybrid precoding is an effective technique employed by mm-Wave systems to increase the spectral efficiency and reduce the cost imposed by massive arrays in a multiple-output multiple-input (MIMO) configuration \cite{mmwaveKeyElements,mimoRHeath}.
	
	
	In recent years, several techniques have been proposed to design the hybrid precoders in mm-Wave MIMO systems. Initial works have focused on the narrow-band scenario \cite{mimoHybridLeus1,mimoRHeath}. However, to effectively utilize the mm-Wave MIMO architectures with relatively larger bandwidth, there are recent, concerted efforts toward developing broadband hybrid beamforming techniques. The key challenge in hybrid beamforming for a broadband frequency-selective channel is designing a common analog beamformer that is shared across all subcarriers while the digital (baseband) beamformer weights need to be specific to a subcarrier. This difference in  hybrid beamforming design of frequency-selective channels from flat-fading case is the primary motivation for considering hybrid beamforming for orthogonal frequency division multiplexing (OFDM) modulation. The optimal beamforming vector in a frequency-selective channel depends on the frequency, i.e., a subcarrier in OFDM, but the analog beamformer in any of the narrow-band hybrid structures cannot vary with frequency. Thus, the common analog beamformer is required to be designed in consideration of impact to all subcarriers, thereby making the hybrid precoding more difficult than the narrow-band case.
	
	In recent studies on broadband systems,  channel estimation is considered in \cite{widebandChannelEst1} and \cite{widebandChannelEst2}, and hybrid beamforming design is investigated in \cite{Alkhateeb_WBHB,sohrabiOFDM,widebandHBWithoutInsFeedback,widebandMLbased} where OFDM-based frequency-selective structures are designed. In particular, \cite{Alkhateeb_WBHB} proposes a Gram-Schmidt orthogonalization based approach for hybrid beamforming (GS-HB) with the assumption of perfect channel state information (CSI) and GS-HB selects the precoders from a finite codebook which are obtained from the instantaneous channel data. In \cite{sohrabiOFDM}, a phase extraction based approach is proposed for hybrid precoder design with the same assumption regarding the CSI. However perfect CSI knowledge is not practical, especially in mm-Wave channel scenario where the channel characteristics change greatly in a short time \cite{coherenceTimeRef}. 

	 The performance of the above works strongly relies on the perfectness of the channel. To relax this dependence and obtain robust performance against the imperfections in the estimated channel matrix, we propose a deep learning (DL) approach for the design of hybrid beamformers in broadband mm-Wave massive MIMO systems. Compared to conventional methods, DL is capable of uncovering complex relationships in data/signals and, thus, can achieve better performance. This has been demonstrated in several successful applications of DL in wireless communications problems such as channel estimation \cite{mimoDLChannelEstimation,deepLearningChannelAndDOAEst}, analog beam selection \cite{mimoDLHybrid,hodge2019multi}, and also hybrid beamforming \cite{mimoDLHybrid,mimoDLChannelModelBeamformingFacebook,mimoDeepPrecoderDesign,elbirDL_COMML,elbirQuantizedCNN2019}. However, these works were proposed for narrow-band scenario \cite{mimoDeepPrecoderDesign,mimoDLChannelModelBeamformingFacebook,elbirDL_COMML,elbirQuantizedCNN2019}. The DL-based design of hybrid precoders for broadband mm-Wave massive MIMO systems, despite its high practical importance, remains unexamined so far.  
	 
	 In this paper, we design a deep convolutional neural network (CNN) to enable hybrid beamformer design in broadband mm-Wave systems. The input of the CNN is the channel matrix of all subcarriers and the output is the hybrid beamformer weights. In a nutshell, the proposed deep network constructs a nonlinear mapping between the channel matrix and the hybrid beamformers. The proposed DL framework has two stages: training (offline) and prediction (online). During training, several channel realizations are generated and hybrid beamforming problem is solved via manifold optimization (MO) approach \cite{hybridBFAltMin} to obtain the network labels. In the prediction stage, when the CNN operates online, the hybrid beamformers are estimated by simply feeding the CNN with the channel matrix.
	 The proposed approach is advantageous since it does not require the perfect channel data in the prediction stage and still provides robust performance. 
	 Moreover, our CNN structure takes less computation time to obtain hybrid beamformers when compared to the conventional approaches. 
	 
	
	The rest of the paper is organized as follows. In the following section, we introduce the system model for mm-Wave channel and formulate the problem. Section~\ref{sec:bb_hb} presents the broadband hybrid beamformer design philosophy. We present our DL framework in Section~\ref{sec:HD_Design}. We follow this with numerical simulations in Section~\ref{sec:Sim} and conclude in Section~\ref{sec:Conc}.
	

	\section{System Model and Problem Formulation}
	\label{sec:SystemModel}
\begin{figure}
		\centering
		\includegraphics[width=1.0\columnwidth]{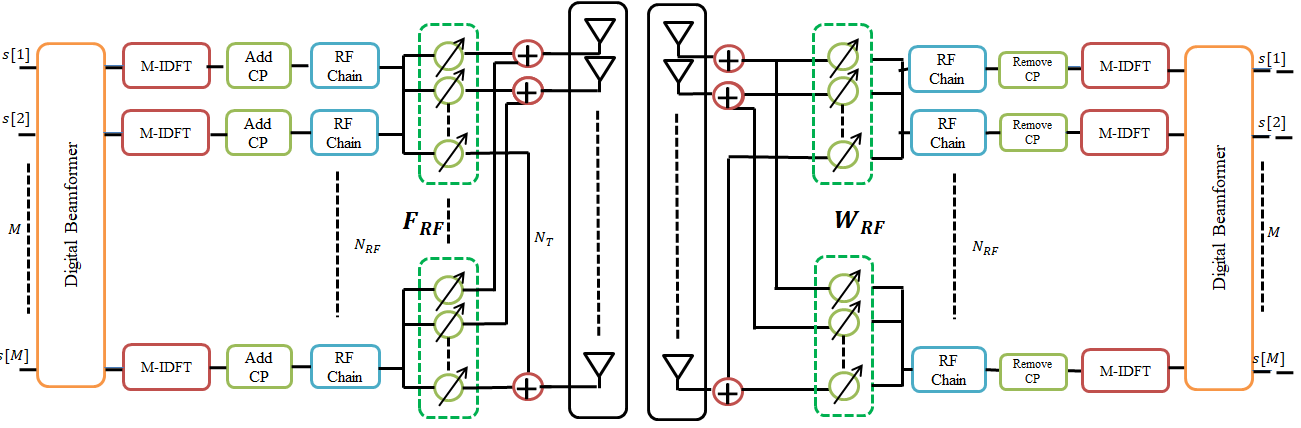} 
		\caption{System architecture of broadband mm-Wave  MIMO-OFDM based transceiver with hybrid  (analog and baseband) beamforming.}
		\label{fig_SystemModel}
	\end{figure}

	We consider the hybrid beamformer design for frequency selective broadband mm-Wave massive MIMO-OFDM system with $M$ subcarriers (Fig.~\ref{fig_SystemModel}).
	The base station (BS) has $N_\mathrm{T}$ antennas and $N_\mathrm{RF}$ $(N_\mathrm{RF} \leq N_\mathrm{T})$ RF chains to transmit $N_\mathrm{S}$ data streams. In the downlink, the BS first precodes $N_\mathrm{S}$ data symbols $\mathbf{s}[m] = [s_1[m],s_2[m],\dots,s_{N_\mathrm{S}}[m]]^\textsf{T}\in \mathbb{C}^{N_\mathrm{S}}$ at each subcarrier by applying the subcarrier-dependent baseband precoders $\mathbf{F}_{\mathrm{BB}}[m] = [\mathbf{f}_{\mathrm{BB}_1}[m],\mathbf{f}_{\mathrm{BB}_2}[m],\dots,\mathbf{f}_{\mathrm{BB}_{N_\mathrm{S}}} [m]]\in \mathbb{C}^{N_{\mathrm{RF}}\times N_\mathrm{S}}$. Then the signal is transformed to the time-domain via $M$-point inverse fast Fourier transforms (IFFTs). After adding the cyclic prefix, the transmitter employs a subcarrier-independent RF precoder $\mathbf{F}_{\mathrm{RF}}\in \mathbb{C}^{N_\mathrm{T}\times N_{\mathrm{RF}}}$ to form the transmitted signal.  Also, given that $\mathbf{F}_{\mathrm{RF}}$ consists of analog phase shifters, we assume that the RF precoder has constant equal-norm elements, i.e., $|[\mathbf{F}_{\mathrm{RF}}]_{i,j}|^2 =1$. In addition, we have the power constraint  $\sum_{m=1}^{M}\|\mathbf{F}_{\mathrm{RF}}\mathbf{F}_{\mathrm{BB}}[m] \|_\mathcal{F}^2= MN_\mathrm{S}$  that is enforced by the normalization of $\mathbf{F}_{\mathrm{BB}}[m] $. Thus, the $N_\mathrm{T}\times 1$ transmitted signal is written as $	\mathbf{x}[m] = \mathbf{F}_{\mathrm{RF}} \mathbf{F}_{\mathrm{BB}}[m]  \mathbf{s}[m].$

	In mm-Wave transmission, the channel is represented by a geometric model with limited scattering \cite{mimoChannelModel1}. Hence, we assume that the channel matrix  $\mathbf{H}[m]$ includes the contributions of $L$ clusters, each of which has the time delay $\tau_l$ and $N_\mathrm{sc} $ scattering paths/rays within the cluster.  Each ray in the $l$th cluster has also a relative time delay $\tau_{{r}}$, AOA/AOD shift $\vartheta_{rl},\varphi_{rl}$ and the complex path gain $\alpha_{l,r}$ for $r = \{1,\dots, N_\mathrm{sc}\}$. Let $p(\tau)$ denote a pulse shaping function for $T_\mathrm{s}$-spaced signaling evaluated  at $\tau$ seconds \cite{channelModelSparseSayeed}, and the mm-Wave delay-$d$ MIMO channel matrix is
	\begin{align}
	\label{eq:delaydChannelModel}
	\mathbf{H}[d] \hspace{-3pt}= \hspace{-3pt} \beta\sum_{l=1}^{L} \sum_{r=1}^{N_\mathrm{sc}}\alpha_{l,r} p(dT_\mathrm{s} \hspace{-3pt}-\hspace{-3pt} \tau_l\hspace{-3pt} - \hspace{-3pt}\tau_{{r}}) 
 \mathbf{a}_\mathrm{R}(\theta_{l}\hspace{-3pt} - \hspace{-3pt}\vartheta_{rl}) \mathbf{a}_\mathrm{T}^\textsf{H}(\phi_l \hspace{-3pt}-\hspace{-3pt} \varphi_{rl}),
	\end{align}
	where $\beta =  \sqrt{\frac{ N_\mathrm{T} N_{\mathrm{R}} } {L}}$ and  $\mathbf{a}_\mathrm{R}(\theta)$, $\mathbf{a}_\mathrm{T}(\phi)$ are the $N_\mathrm{R} \times 1$ and $N_\mathrm{T}\times 1$ steering vectors representing the array responses of the receive and transmit antenna arrays respectively. Note that the angular parameters $\theta$ and $\phi$ correspond to the elevation and the azimuth angles, respectively.
Then for a uniform linear array (ULA), the array response of the transmit array is $\mathbf{a}_\mathrm{T}(\phi) = \big[ 1, e^{j\frac{2\pi}{\lambda} \bar{d}_\mathrm{T}\sin(\phi)},\dots,e^{j\frac{2\pi}{\lambda} (N_\mathrm{T}-1)\bar{d}_\mathrm{T}\sin(\phi)} \big]^\textsf{T},$
	where $\bar{d}_\mathrm{T}=\bar{d}_\mathrm{R} = \lambda/2$ is the antenna spacing and  $\mathbf{a}_\mathrm{R}(\theta)$ is defined in a similar way as for $\mathbf{a}_\mathrm{T}(\phi)$.	By using the delay-$d$ channel model in (\ref{eq:delaydChannelModel}), the channel matrix at subcarrier $m$ is expressed as $	\mathbf{H}[m] = \sum_{d=0}^{D-1}\mathbf{H}[d]e^{-j\frac{2\pi m}{M} d},$
	where $D$ is the length of cyclic prefix \cite{channelModelSparseBajwa}.

		Assuming a block-fading channel  model \cite{Alkhateeb_WBHB}, the received signal at subcarrier $m$ is
	\begin{align}
	\label{arrayOutput}
	\mathbf{y}[m] = \sqrt{\rho}\mathbf{H}[m] \mathbf{F}_\mathrm{RF}\mathbf{F}_\mathrm{BB}[m]\mathbf{s}[m] + \mathbf{n}[m],
	\end{align}
	where $\rho$ represents the average received power and  $\mathbf{H}[m]\in \mathbb{C}^{N_\mathrm{R}\times N_\mathrm{T}}$ channel matrix and $\mathbf{n}[m] \sim \mathcal{CN}(\mathbf{0},\sigma^2 \mathbf{I}_\mathrm{N_\mathrm{R}})$ is  additive white Gaussian noise (AWGN) vector.

	At the receiver, the received signal is first processed by analog combiner $\mathbf{W}_\mathrm{RF}$, then the cyclic prefix is removed from the the processed signal and $N_\mathrm{RF}$ $M$-point FFTs are applied to recover the signal in frequency domain. Finally, the receiver employs low-dimensional $N_\mathrm{R}\times N_\mathrm{S}$ digital combiners $\{\mathbf{W}_\mathrm{BB}[m]\}_{m\in \mathcal{M}}$ where $\mathcal{M} = \{1,\dots,M\}$. The received and processed signal is obtained as $\tilde{\mathbf{y}}[m] =  \sqrt{\rho}\mathbf{W}_\mathrm{BB}^\textsf{H}[m]\mathbf{W}_\mathrm{RF}^\textsf{H}\mathbf{H}[m] \mathbf{F}_\mathrm{RF}\mathbf{F}_\mathrm{BB}[m]\mathbf{s}[m] 
	+ \mathbf{W}_\mathrm{BB}^\textsf{H}[m]\mathbf{W}_\mathrm{RF}^\textsf{H}\mathbf{n}[m],$
	where the analog combiner $\mathbf{W}_\mathrm{RF}\in \mathbb{C}^{N_\mathrm{R}\times N_\mathrm{RF}}$ has the constraint $\big[[\mathbf{W}_\mathrm{RF}]_{:,i}[\mathbf{W}_\mathrm{RF}]_{:,i}^\textsf{H}\big]_{i,i}=1$ similar to the RF precoder. 
	In a practical scenario,  the estimated channel matrix is obtained by channel estimation techniques \cite{channelEstLargeArrays,channelEstimation1}. We assume the channel matrix is available only in the training stage of our DL framework to obtain the labels (e.g., hybrid beamformers) and, in the prediction stage, the proposed CNN does not require the perfect channel data. 
	
	In this paper, we focus on the design of hybrid precoders $\mathbf{F}_\mathrm{RF},\mathbf{F}_\mathrm{BB}[m]$, $\mathbf{W}_\mathrm{RF},\mathbf{W}_\mathrm{BB}[m]$ by maximizing the overall spectral efficiency of the system under power spectral density constraint for each subcarrier. We use the spectral efficiency as a performance metric for the hybrid beamformer design problem as in the most of the earlier works using spectral efficiency or mutual information \cite{mimoHybridLeus1,Alkhateeb_WBHB} while there are also studies designing the beamformers by optimizing the bit error rate (BER). The hybrid beamformer design problem is
	\begin{align}
	\label{HBdesignProblem}
	&\underset{\mathbf{{F}}_\mathrm{RF},\mathbf{{W}}_\mathrm{RF}, \{\mathbf{{F}}_\mathrm{BB}[m],\mathbf{{W}}_\mathrm{BB}[m]\}_{m\in \mathcal{M}}}{\maximize} \;\;\;\frac{1}{M}\sum_{m =1}^{M} R[m] \nonumber \\
	&\subjectto \;\;\;\; \mathbf{{F}}_\mathrm{RF} \in \mathcal{F}_\mathrm{RF},   \mathbf{{W}}_\mathrm{RF} \in \mathcal{W}_\mathrm{RF}, \nonumber \\
	&\;\;\;\;\;\;\;\;\;\;\;\;\;\;\;\;\;\;\;\;\sum_{m=1}^{M}||\mathbf{{F}}_\mathrm{RF}\mathbf{{F}}_\mathrm{BB}[m]||_{\mathcal{F}}^2 = M N_\mathrm{S},
	\end{align}
	where $\mathcal{F}_\mathrm{RF}$ and $\mathcal{W}_\mathrm{RF}$ are the feasible sets for the RF precoder and combiners which obey the unit-norm constraint \cite{Alkhateeb_WBHB,hybridBFAltMin} and $R[m]$ is the overall spectral efficiency of the subcarrier $m$. Assuming that the Gaussian symbols are transmitted through the mm-Wave channel \cite{mimoRHeath,mimoHybridLeus1}, we have $R[m]=\textrm{log}_2 \bigg| \mathbf{I}_{N_\mathrm{S}} 
	+\frac{\rho}{N_\mathrm{S}}\boldsymbol{\Lambda}_\mathrm{n}^{-1}\mathbf{W}_\mathrm{BB}^\textsf{H}\mathbf{W}_\mathrm{RF}^\textsf{H}  \mathbf{H}[m] \times\mathbf{{F}}_\mathrm{RF}\mathbf{{F}}_\mathrm{BB}[m]\mathbf{{F}}_\mathrm{BB}^\textsf{H}[m] \mathbf{{F}}_\mathrm{RF}^\textsf{H}\mathbf{H}^\textsf{H}[m]\mathbf{W}_\mathrm{RF}\mathbf{W}_\mathrm{BB}[m] \bigg|,$
	where $\boldsymbol{\Lambda}_\mathrm{n} = \sigma_n^2 \mathbf{W}_\mathrm{BB}^\textsf{H}[m]\mathbf{W}_\mathrm{RF}^\textsf{H} \mathbf{W}_\mathrm{RF}\mathbf{W}_\mathrm{BB}[m]\in \mathbb{C}^{N_\mathrm{S} \times N_\mathrm{S}}$ corresponds to the covariance of the noise term of the received signal.

	
	Our goal is to design a deep neural network to estimate the hybrid beamformers $\mathbf{F}_\mathrm{RF},\mathbf{F}_\mathrm{BB}[m]$ and $\mathbf{W}_\mathrm{RF},\mathbf{W}_\mathrm{BB}[m]$ by using the available imperfect channel matrix $\{\mathbf{H}[m]\}_{m\in \mathcal{M}}$ in the sense that maximum spectral efficiency is achieved.
	
	\section{Broadband Hybrid Beamformer Design}
	\label{sec:bb_hb}
	The design problem of the hybrid beamformers requires a joint optimization as in (\ref{HBdesignProblem}), however this approach is computationally complex and even intractable. Instead, a decoupled problem is preferred \cite{mimoRHeath,sohrabiOFDM,elbirQuantizedCNN2019,hybridBFAltMin}. In this respect, first the hybrid precoders $\mathbf{F}_\mathrm{RF},\mathbf{F}_\mathrm{BB}[m]$ are estimated, then the hybrid combiners $\mathbf{W}_\mathrm{RF},\mathbf{W}_\mathrm{BB}[m]$ are found. Following this approach, the hybrid precoder design problem is written as the maximization of the mutual information achieved by Gaussian signaling over mm-Wave channel \cite{Alkhateeb_WBHB}, i.e.,
	\begin{align}
	\label{PrecoderDesignProblem}
	&\underset{\mathbf{{F}}_\mathrm{RF}, \{\mathbf{{F}}_\mathrm{BB}[m]\}_{m\in \mathcal{M}}}{\maximize} \frac{1}{M}\sum_{m =1}^{M} \mathcal{I}\{\mathbf{F}_\mathrm{RF},\mathbf{F}_\mathrm{BB}[m]\} \nonumber \\
	&\subjectto \; \mathbf{{F}}_\mathrm{RF} \in \mathcal{F}_\mathrm{RF}, 
	\sum_{m=1}^{M}||\mathbf{{F}}_\mathrm{RF}\mathbf{{F}}_\mathrm{BB}[m]||_{\mathcal{F}}^2 = M N_\mathrm{S},
	\end{align}
	where $\mathcal{I}\{\mathbf{F}_\mathrm{RF},\mathbf{F}_\mathrm{BB}[m]\} = \textrm{log}_2 \bigg| \mathbf{I}_{N_\mathrm{S}}  + \boldsymbol{\Gamma}[m]\bigg|$ and $\boldsymbol{\Gamma}[m] = \frac{\rho}{N_\mathrm{S}}\mathbf{H}[m]\mathbf{{F}}_\mathrm{RF}\mathbf{{F}}_\mathrm{BB}[m]\mathbf{{F}}_\mathrm{BB}^\textsf{H}[m] \mathbf{{F}}_\mathrm{RF}^\textsf{H}\mathbf{H}^\textsf{H}[m]$.
	We note here that the optimization problem in (\ref{PrecoderDesignProblem}) is approximated by using similarity between the hybrid beamformer $\mathbf{F}_\mathrm{RF}\mathbf{F}_\mathrm{BB}[m]$ and the optimal unconstrained beamformer $\mathbf{F}^{\mathrm{opt}}[m]$ which is obtained from the right singular matrix of the channel matrix $\mathbf{H}[m]$ \cite{hybridBFAltMin,mimoRHeath}. In particular, we can denote the singular value decomposition of the channel matrix as  $\mathbf{H}[m] = \mathbf{U}[m] \boldsymbol{\Sigma}[m] \mathbf{V}^H[m]$, where $\mathbf{U}[m]\in \mathbb{C}^{N_\mathrm{R}\times \mathrm{rank}(\mathbf{H}[m])}$ and $\mathbf{V}[m]\in \mathbb{C}^{N_\mathrm{T} \times \mathrm{rank}(\mathbf{H}[m])}$ are the left and the right singular value matrices of the channel matrix, respectively, 
	and $\boldsymbol{\Sigma}[m]$ is $\mathrm{rank}(\mathbf{H}[m])\times \mathrm{rank}(\mathbf{H}[m])$ matrix composed of the singular values of $\mathbf{H}[m]$ in descending order. By decomposing $\boldsymbol{\Sigma}[m]$ and $\mathbf{V}[m]$ as $\boldsymbol{\Sigma}[m] = \mathrm{diag}\{ \tilde{\boldsymbol{\Sigma}}[m],\bar{\boldsymbol{\Sigma}}[m] \},\hspace{5pt} \mathbf{V}[m] = [\tilde{\mathbf{V}}[m],\bar{\mathbf{V}}[m]],$
	where $\tilde{\mathbf{V}}[m]\in \mathbb{C}^{N_\mathrm{T}\times N_\mathrm{S}}$, one can readily select the unconstrained precoder as $\mathbf{F}^{\mathrm{opt}}[m] = \tilde{\mathbf{V}}[m]$ \cite{mimoRHeath}. Then, hybrid precoder design problem for subcarrier $m$ is the minimization of the Euclidean distance between $\mathbf{F}^{\mathrm{opt}}[m]$ and $\mathbf{F}_\mathrm{RF}\mathbf{F}_\mathrm{BB}[m]$:
	\begin{align}
	\label{PrecoderSingleCarrier}
	&\underset{\mathbf{F}_\mathrm{RF},\mathbf{F}_\mathrm{BB}[m]}{\minimize} \big|\big|   \mathbf{F}^{\mathrm{opt}}[m]  - \mathbf{F}_\mathrm{RF}\mathbf{F}_\mathrm{BB}[m]  \big|\big|_\mathcal{F}^2
	\nonumber \\
	&\subjectto \; \mathbf{{F}}_\mathrm{RF} \in \mathcal{F}_\mathrm{RF}, 
	\big|\big| \mathbf{{F}}_\mathrm{RF}\mathbf{{F}}_\mathrm{BB}[m]\big|\big|_{\mathcal{F}}^2 =  N_\mathrm{S}.
	\end{align}
	To incorporate all subcarriers, we rewrite (\ref{PrecoderSingleCarrier}) as
	\begin{align}
	\label{PrecoderAllCarriers}
	&\underset{\mathbf{F}_\mathrm{RF},\{\mathbf{F}_\mathrm{BB}[m]\}_{m \in \mathcal{M}}}{\minimize} \big|\big|   \tilde{\mathbf{F}}^{\mathrm{opt}}  - \mathbf{F}_\mathrm{RF}\tilde{\mathbf{F}}_\mathrm{BB}  \big|\big|_\mathcal{F}^2
	\nonumber \\
	&\subjectto \; \mathbf{{F}}_\mathrm{RF} \in \mathcal{F}_\mathrm{RF}, 
	\sum_{m=1}^{M}\big|\big| \mathbf{{F}}_\mathrm{RF}\mathbf{{F}}_\mathrm{BB}[m]\big|\big|_{\mathcal{F}}^2 =  MN_\mathrm{S},
	\end{align}
	where $\tilde{\mathbf{F}}^{\mathrm{opt}}\in \mathbb{C}^{N_\mathrm{T}\times MN_\mathrm{S}}$ and $\tilde{\mathbf{F}}_\mathrm{BB} \in \mathbb{C}^{N_\mathrm{RF}\times MN_\mathrm{S}} $ contain the beamformers for all subcarriers and they are given respectively as $	\tilde{\mathbf{F}}^{\mathrm{opt}} = \big[ \mathbf{F}^{\mathrm{opt}}[1],\mathbf{F}^{\mathrm{opt}}[2],\dots,\mathbf{F}^{\mathrm{opt}}[M] \big],$  
	and $\tilde{\mathbf{F}}_\mathrm{BB} = \big[ \mathbf{{F}}_\mathrm{BB}[1],\mathbf{{F}}_\mathrm{BB}[2],\dots,\mathbf{{F}}_\mathrm{BB}[M]\big].$

	Once the hybrid precoders are designed, the hybrid combiners $\mathbf{W}_\mathrm{RF},\mathbf{W}_\mathrm{BB}[m]$ are determined by minimizing the mean-square-error (MSE), $\mathbb{E}\{\big|\big| \mathbf{s}[m] - \mathbf{W}_\mathrm{BB}^\textsf{H}[m] \mathbf{W}_\mathrm{RF}^\textsf{H}\mathbf{y}[m]  \big|\big|_2^2\},$. The combiner-only design problem is
	\begin{align}
	\label{CombinerOnlyProblem}
	&\underset{\mathbf{W}_\mathrm{RF}, \mathbf{W}_\mathrm{BB}[m] }{\minimize}
	\mathbb{E}\{\big|\big| \mathbf{s}[m] - \mathbf{W}_\mathrm{BB}^\textsf{H}[m] \mathbf{W}_\mathrm{RF}^\textsf{H}\mathbf{y}[m]  \big|\big|_2^2\} \nonumber \\
	&\subjectto \mathbf{W}_\mathrm{RF} \in{\mathcal{W}}_\mathrm{RF}.
	\end{align}
A more efficient form of the problem in (\ref{CombinerOnlyProblem}) is due to \cite{mimoRHeath}, of which we skip the details due to paucity of space here.
	Similar to the precoder design in (\ref{PrecoderSingleCarrier}) and (\ref{PrecoderAllCarriers}), we solve the combiner design problem in (\ref{CombinerOnlyProblem}) for all subcarriers. We first define the MMSE combiner $\mathbf{W}_\mathrm{MMSE}^\textsf{H}[m] = \frac{1}{\rho}\big( \mathbf{F}^{\mathrm{opt}^\textsf{H}}[m]\mathbf{H}^\textsf{H}[m]\mathbf{H}[m]\mathbf{F}^{\mathrm{opt}}[m]+ \frac{N_\mathrm{S}\sigma_n^2}{\rho}\mathbf{I}_{N_\mathrm{S}} \big)^{-1}  \mathbf{F}^{\mathrm{opt}^\textsf{H}}[m]\mathbf{H}^\textsf{H}[m]$. Then, we construct  $\tilde{\mathbf{W}}_\mathrm{MMSE}\hspace{-2pt} = \hspace{-3pt} \big[{\mathbf{W}}_\mathrm{MMSE}[1],{\mathbf{W}}_\mathrm{MMSE}[2],\dots,{\mathbf{W}}_\mathrm{MMSE}[M] \big],$
	and 
	\begin{align}
	    	\tilde{\mathbf{W}}_\mathrm{BB} = \big[{\mathbf{W}}_\mathrm{BB}[1],{\mathbf{W}}_\mathrm{BB}[2],\dots,{\mathbf{W}}_\mathrm{BB}[M]  \big].
	\end{align}
	Thus, the hybrid combiner design problem is
	\begin{align}
	\label{CombinerOnlyProblemAllSubcarriers}
	&\underset{\mathbf{W}_\mathrm{RF}, \{\mathbf{W}_\mathrm{BB}[m]\}_{m\in \mathcal{M}}}{\minimize}
	\big|\big| \tilde{\mathbf{W}}_\mathrm{MMSE}
	- \mathbf{W}_\mathrm{RF} \tilde{\mathbf{W}}_\mathrm{BB}\big|\big|_\mathcal{F}^2 \nonumber \\
	&\subjectto \;\;\;\;\;\mathbf{W}_\mathrm{RF} \in{\mathcal{W}}_\mathrm{RF} \nonumber \\
	&\;\;\;\;\;\;\;\;\;\;\;\;\;\;\;\;\;\;\;\;\;\;\; \mathbf{W}_\mathrm{BB}[m] = (\mathbf{W}_\mathrm{RF}^\textsf{H} \boldsymbol{\Lambda}_\mathrm{y}[m]  \mathbf{W}_\mathrm{RF})^{-1}\nonumber \\
	&\;\;\;\;\;\;\;\;\;\;\;\;\;\;\;\;\;\;\;\;\;\;\;\times (\mathbf{W}_\mathrm{RF}^\textsf{H}\boldsymbol{\Lambda}_\mathrm{y}[m]\mathbf{W}_\mathrm{MMSE}[m],
	\end{align}
	where $\boldsymbol{\Lambda}_\mathrm{y}[m] = 
		\rho\mathbf{H}[m]\mathbf{F}_\mathrm{RF}\mathbf{F}_\mathrm{BB}[m]\mathbf{F}_\mathrm{BB}^\textsf{H}[m]\mathbf{F}_\mathrm{RF}^\textsf{H}\mathbf{H}^\textsf{H}[m] + \sigma_n^2\mathbf{I}_{N_\mathrm{R}}$ denotes the covariance of the array output in (\ref{arrayOutput}).
	
	The optimization problems in (\ref{PrecoderAllCarriers}) and (\ref{CombinerOnlyProblemAllSubcarriers}) do not require a codebook which is the set of array response of the transmit and receive arrays \cite{mimoRHeath}. In fact, the manifold optimization problem for  (\ref{PrecoderAllCarriers}) and (\ref{CombinerOnlyProblemAllSubcarriers}) is initialized from a random point, i.e., beamformers with unit-norm and random phases.
	


	\vspace{-14pt}
	\section{Learning-Based Hybrid Beamformer Design}
	\label{sec:HD_Design}
	We now design a CNN which accepts the channel matrix as input.  Once the deep network is fed with the channel matrix of all subcarriers, the hybrid beamformers is obtained at the output of the CNN, which is a regression layer. In order to obtain the labels of the network (i.e., $\mathbf{F}_\mathrm{RF},\mathbf{F}_\mathrm{BB}[m],\mathbf{W}_\mathrm{RF},\mathbf{W}_\mathrm{BB}[m]$), we solve the hybrid beamformer design problem stated in (\ref{PrecoderAllCarriers}) and (\ref{CombinerOnlyProblemAllSubcarriers}). Then the training data is generated in an offline process. The algorithmic steps for training data generation is presented in Algorithm~\ref{alg:algorithmTraining}.
	\begin{algorithm}[H]
		\begin{algorithmic}[1]
			\caption{Training data generation for DLHB. }
			\Statex {\textbf{Input:} $N$,  $G$, $M$,  SNR$_{\text{TRAIN}}$}.  {\textbf{Output:}  $\mathcal{D}_{\text{TRAIN}}$.}
			\label{alg:algorithmTraining}
			\State Generate  $\{\mathbf{H}^{(n)}[m]\}_{n=1}^N$  for $m \in \mathcal{M}$.
			\State Initialize with $t=1$ while the dataset length is $T=NG$.
			\State   \textbf{for}  $1 \leq n \leq N$ \textbf{and}  $1 \leq g \leq G$ \textbf{do}
			\State \indent  $[\tilde{\mathbf{H}}^{(n,g)}[m]]_{i,j} \sim \mathcal{CN}([\mathbf{H}^{(n)}[m]]_{i,j},\sigma_{\text{TRAIN}}^2)$.
			\State \indent  Using $\mathbf{H}^{(n,g)}[m]$, find  $\hat{\mathbf{F}}_{\mathrm{RF}}^{(n,g)}$ and $\hat{\mathbf{F}}_{\mathrm{BB}}^{(n,g)}[m]$ \par \indent   by solving  (\ref{PrecoderAllCarriers}).
			\State \indent  Find  $\hat{\mathbf{W}}_{\mathrm{RF}}^{(n,g)}$ and $\hat{\mathbf{W}}_{\mathrm{BB}}^{(n,g)}[m]$   by solving  (\ref{CombinerOnlyProblemAllSubcarriers}).
			\State \indent \textbf{for}  $1 \leq m \leq M$ \textbf{do}
			\State   \indent \indent$[[\mathbf{X}^{(t)}[m]]_{:,:,1}]_{i,j} = |[\tilde{\mathbf{H}}^{(n,g)}[m]]_{i,j}|$.
			\State  \indent\indent $[[\mathbf{X}^{(t)}[m]]_{:,:,2}]_{i,j}=\operatorname{Re} \{[\tilde{\mathbf{H}}^{(n,g)}[m]]_{i,j}\}$ .
			\State  \indent\indent $[[\mathbf{X}^{(t)}[m]]_{:,:,3}]_{i,j} = \operatorname{Im}\{[\tilde{\mathbf{H}}^{(n,g)}[m]]_{i,j}\}$ $\forall ij$.
			\State \indent\textbf{end for} $m$,
			\State \indent Construct $\mathbf{X}^{(t)} \hspace{-2pt}=  [\mathbf{X}^{(t)^\textsf{T}}[1],\dots,\mathbf{X}^{(t)^\textsf{T}}[M] ]^\textsf{T}$, and  $\mathbf{z}^{(t)}$.
			\State \indent $t = t+1$.
			
			\State \indent\textbf{end for} $g,n$,	
			\State $\mathcal{D}_{\text{TRAIN}} = \big((\mathbf{X}^{(1)}, \mathbf{z}^{(1)}),(\mathbf{X}^{(2)}, \mathbf{z}^{(2)}),\dots, (\mathbf{X}^{(T)}, \mathbf{z}^{(T)})\big).$
		\end{algorithmic}
	\end{algorithm}

	In order to form the input of the deep network, the channel matrix is partitioned to three components to feed the network with different features \cite{elbirDL_COMML}. Hence we use real, imaginary parts and  the absolute value of each entry of the channel matrix respectively.  This approach provides good features for the solution of the DL problem \cite{elbirQuantizedCNN2019,elbirDL_COMML,elbirIETRSN2019}. Let $\mathbf{X} = [\mathbf{X}^\textsf{T}[1],\dots,\mathbf{X}^\textsf{T}[M] ]^\textsf{T}\in \mathbb{R}^{ MN_\mathrm{R} \times N_\mathrm{T}\times 3}$ be the real-valued input of the network as a three dimensional matrix. Then, for a channel matrix $\mathbf{H}[m]\in \mathbb{C}^{N_\mathrm{R}\times N_\mathrm{T}}$,  the $(i,j)$-th entry of the submatrices per subcarrier is  $[[\mathbf{X}[m]]_{:,:,1}]_{i,j} = | [\mathbf{H}[m]]_{i,j}| $ for the first "channel" of $\mathbf{X}[m]$. The second and the third "channels" are given by $[[\mathbf{X}[m]]_{:,:,2}]_{i,j} = \operatorname{Re}\{[\mathbf{H}[m]]_{i,j}\}$ and $[[\mathbf{X}[m]]_{:,:,3}]_{i,j} = \operatorname{Im}\{[\mathbf{H}[m]]_{i,j}\}$, respectively. 
	
	The output layer of the network is a regression layer which contains the hybrid precoder and combiners. Hence, the output layer is the vectorized form of analog beamformers and baseband beamformers for all subcarriers. We can write the network output $\mathbf{z} \in \mathbb{R}^{N_\mathrm{S}\big(N_\mathrm{T} + N_\mathrm{R} + 4MN_\mathrm{RF} \big) \times 1}$ as $\mathbf{z} = [ \mathbf{z}_\mathrm{RF}^\textsf{T},\;\; \tilde{\mathbf{z}}_\mathrm{BB}^\textsf{T} ]^\textsf{T},$
	where $\mathbf{z}_\mathrm{RF} = [\mathrm{vec}\{\angle \mathbf{F}_\mathrm{RF}  \}^\textsf{T},\mathrm{vec}\{\angle \mathbf{W}_\mathrm{RF}  \}^\textsf{T}]^\textsf{T}$ is a real-valued $N_\mathrm{S}(N_\mathrm{T} + N_\mathrm{R})\times 1$ vector and $ \tilde{\mathbf{z}}_\mathrm{BB} = [\mathbf{z}_\mathrm{BB}^\textsf{T}[1],\mathbf{z}_\mathrm{BB}^\textsf{T}[2],\dots,\mathbf{z}_\mathrm{BB}^\textsf{T}[M]]^\textsf{T} \in \mathbb{R}^{4M N_\mathrm{S} N_\mathrm{RF}}$  includes the baseband beamformers for all subcarriers where 
	\begin{align}
	&\mathbf{z}_\mathrm{BB}[m] = [\mathrm{vec}\{\operatorname{Re}\{ \mathbf{F}_\mathrm{BB}[m]\} \}^\textsf{T}, \mathrm{vec}\{\operatorname{Im}\{ \mathbf{F}_\mathrm{BB}[m]\} \}^\textsf{T}, \nonumber \\
	&\;\;\;\;\;\;\;\;\;\;\;\mathrm{vec}\{\operatorname{Re}\{ \mathbf{W}_\mathrm{BB}[m]\} \}^\textsf{T}, \mathrm{vec}\{\operatorname{Im}\{ \mathbf{W}_\mathrm{BB}[m]\} \}^\textsf{T}]^\textsf{T}.
	\end{align}

\begin{figure}[t]
		\centering
		{\includegraphics[draft=false,scale=0.4]{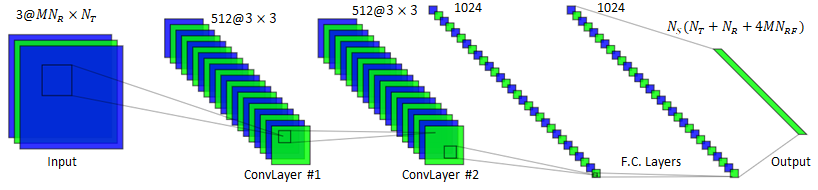} }
		\caption{Proposed network architecture.\vspace{-18pt}}
		\label{fig_Network}
	\end{figure}

	The proposed network is composed of ten layers as  shown in Fig.~\ref{fig_Network}. The first layer is the input layer accepting the channel matrix data of size $ MN_\mathrm{R}\times N_\mathrm{T}\times 3$. The second and the fourth layer are the convolutional layers with 512 filters of size $3\times 3$ to extract the features hidden in the input data. After each convolutional layer, there is a normalization layer to normalize the output and provide better convergence. The sixth and eighth layers are fully connected layers with 1024 units, respectively. There are dropout layers after the fully connected layers (the seventh and ninth layers) with a 50\% probability. The output layer is the regression layer  with $N_\mathrm{S}\big(N_\mathrm{T} + N_\mathrm{R} + 4MN_\mathrm{RF} \big) $ units which include the hybrid beamformers. The current network parameters are obtained from a hyperparameter tuning process providing the best performance for the considered scenario \cite{elbirDL_COMML,elbirQuantizedCNN2019,elbirIETRSN2019}. 
	
	The proposed deep network is realized and trained in MATLAB on a PC with a single GPU and a  768-core processor.  We have used the stochastic gradient descent algorithm with momentum 0.9  and  updated the network parameters with learning rate $0.005$ and mini-batch size of $128$ samples for 100 epochs. To train the proposed CNN structure, $N=500$ different scenarios are realized as in Algorithm~\ref{alg:algorithmTraining}. For each channel matrix, AWGN is added for different powers of SNR$_{\text{TRAIN}}\in \{15,20,25\}$dB with $G=100$ to account for different channel characteristics. The use of multiple SNR$_{\text{TRAIN}}$ levels provides a wide range of corrupted data in the training which improves the learning and robustness of the network \cite{elbirDL_COMML,elbirQuantizedCNN2019}. Hence, the total size of the training data is $MN_\mathrm{R}\times N_\mathrm{T}\times 3 \times 15000$. In the training process, $80\%$ and $20\%$ of all generated data  are selected as the training and validation datasets, respectively. The validation data is used to test the performance of the network in the simulations for $J_\mathrm{T}=100$ Monte Carlo trials. In order to resemble the corruption in the channel data we also add synthetic noise to the test data where the SNR during testing is defined similar to SNR$_\text{TRAIN}$ as SNR$_\text{TEST} = 20\log_{10}(\frac{|[\mathbf{H}]_{i,j}|^2}{\sigma_{\text{TEST}}^2})$. 
	
    
    \vspace{-14pt}
	\section{Numerical Experiments}
	\label{sec:Sim}
	We evaluated the performance of the proposed DL framework through several experiments. We compared our DL-based hybrid beamforming approach (henceforth called DLHB) with the state-of-the-art hybrid beamforming such as  GS-HB \cite{Alkhateeb_WBHB}, the method in \cite{sohrabiOFDM} (hereafter, \textit{Sohrabi et al.}) and the fully digital beamformer. We also present the performance of manifold optimization (MO) algorithm \cite{hybridBFAltMin} which is used when obtaining the labels of the CNN. To train the deep network, we follow the steps from the previous section with $N_\mathrm{T}=256$ and $N_\mathrm{R}=16$ antennas and $N_\mathrm{RF} = N_\mathrm{S}=4$. We select $M=16$ subcarriers at $f_c = 60$ GHz with $4$ GHz bandwidth, and  $L=10$ clusters  with $N_\mathrm{sc}=5$ scatterers. 
	
\begin{figure}[t]
		\centering
		{\includegraphics[width=0.78\columnwidth]{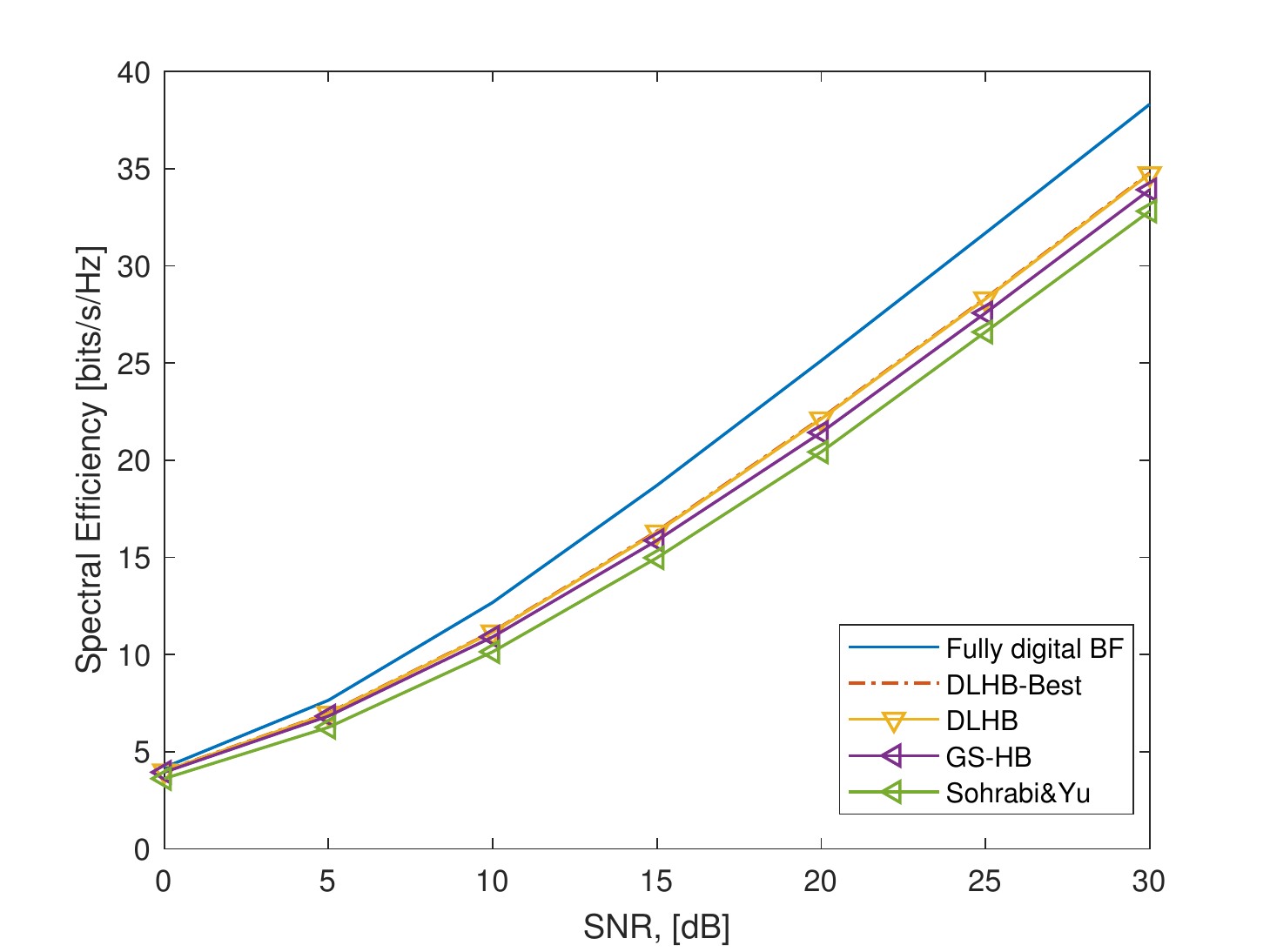} } 
		\caption{Spectral efficiency versus SNR for $N_\mathrm{T}=256$, $N_\mathrm{R}=16$, $N_\mathrm{RF} = N_\mathrm{S}=4$, $M=16$ subcarriers, $L=10$ clusters, and $N_\mathrm{sc}=5$.\vspace{-14pt}}
		\label{fig_SNR_TEST}
	\end{figure}
\begin{figure}[t]
		\centering
		{\includegraphics[width=0.78\columnwidth]{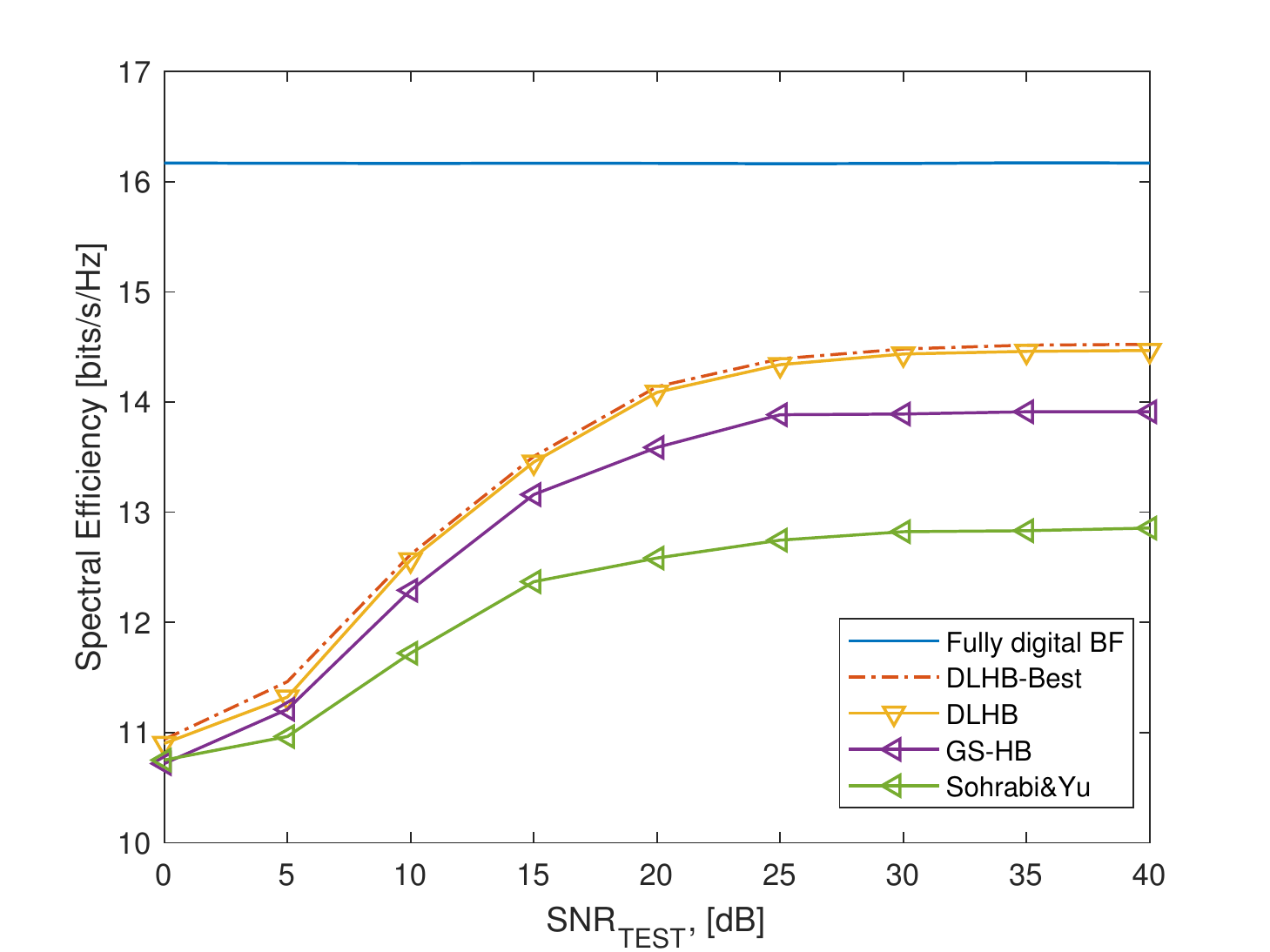} } 
		\caption{Spectral efficiency versus SNR$_\mathrm{TEST}$. SNR$=20$ dB. 
\vspace{-18pt}}
		\label{fig_Corruption_TEST}
	\end{figure}

	Figure~\ref{fig_SNR_TEST} shows the performance of the algorithms with respect to SNR. We can see that DLHB outperforms the competing algorithms and closely follows the MO algorithm. Note that the performance of DLHB is upper bounded by the performance of MO algorithm since MO is involved in the labeling process of our DL framework. In Fig.~\ref{fig_Corruption_TEST}, we investigate the robustness of the algorithms against the imperfect channel input. The enriched training datacorrupted by AWGN to represent the channel imperfections leads to a more robust DLHB performance. The effectiveness of DLHB is attributed to the selection of ``best'' hybrid beamformer vectors via MO algorithm and the feature extraction capability inherit in the input channel data. 
	
	We also compared the computation times of the competing algorithms for the same simulation settings and obtain that DLHB takes about $0.003$ s to predict the hybrid precoders whereas MO, GS-HB and \textit{Sohrabi et al.} need approximately $4.115$ s, $0.006$ s and $0.007$ s respectively. The complexity of DLHB is the multiplication of the input data by the network weights whereas the other algorithms involve optimization \cite{hybridBFAltMin}, greedy search \cite{Alkhateeb_WBHB} and phase extraction \cite{sohrabiOFDM}.

%
%
    \vspace{-16pt}
	\section{Summary}
	\label{sec:Conc}
	We introduced a DL framework for the hybrid beamformer design problem in broadband mm-Wave MIMO systems. The proposed approach has superior performance against the conventional techniques in terms of spectral efficiency and robustness against the imperfect channel data. The proposed DL approach also enjoys less computation time to estimate the hybrid beamformers. 
	
	\bibliographystyle{ieeetr}
	\bibliography{main}

\end{document}